\documentstyle[graphicx]{elsart}
\newcommand{\im}{\mathop{\mathrm{Im}}}

\begin{document}
\begin{frontmatter}

\title{STRUCTURE AND DECAY PROPERTIES OF SPIN--DIPOLE GIANT RESONANCES
WITHIN \\A SEMIMICROSCOPICAL APPROACH}
\author[Moscow]{E.A.~Moukhai},
\author[Moscow]{V.A.~Rodin},
\author[Moscow,Groningen]{M.H.~Urin \thanksref{e-mail}}
\address[Moscow]{Department of Theoretical Nuclear Physics,
Moscow Engineering Physics Institute, Moscow, 115409, Russia}
\address[Groningen]{Kernfysisch Versneller Institute, 9747AA Groningen,
The Netherlands}
\thanks[e-mail]{e-mail: urin@theor.mephi.msk.su, urin@kvi.nl}

\begin{abstract}
A semimicroscopical approach
is applied to calculate: (i) strength functions for the charge-exchange
spin-dipole giant resonances in the \nuc{208}{Pb} parent nucleus; (ii)
partial
and total branching ratios for the direct proton decay of the resonance
in
\nuc{208}{Bi}. The approach is based on continuum-RPA calculations of
corresponding
reaction-amplitudes and phenomenological description of the
doorway-state coupling to many-quasiparticle configurations. The
only adjustable parameter needed for the description is found by
comparison of the calculated and experimental
total widths of the resonance. Other model parameters used in
calculations are
taken from independent data. The calculated total branching
ratio is found to be in reasonable
agreement with the experimental value.
\end{abstract}
\end{frontmatter}
\pagebreak

\section{Introduction}

The intensive experimental and theoretical studies of the direct
nucleon decay of various giant resonances have been  undertaken in 
recent years in an attempt to understand better the interplay of 
single-quasiparticle, collective and many-quasiparticle modes of 
nuclear motion. This work was stimulated by the appearance of 
experimental data on (total) branching ratio
for the proton decay of the spin-dipole giant resonance of the
$(\mathrm{pn}^{\mathrm{-1}})$ - type
($SDR^{(-)}$) in \nuc{208}{Bi}~\cite{akimun}. The branching ratio
was deduced from the analysis of the
\nuc{208}{Pb}(\nuc3{He},t)\nuc{208}{Bi} and
\nuc{208}{Pb}(\nuc3{He},tp)\nuc{207}{Pb} reactions cross sections at
E(\nuc3{He})=450~MeV.
Being related to the spin-flip giant resonance, this branching ratio is a 
valuable addition to the experimental data on partial proton widths 
of the Gamow-Teller resonance ($GTR$), which were also obtained in
Ref.~\cite{akimun}.

The main aim of this work is to calculate the above-mentioned branching
ratio within the approach proposed previously~\cite{murur,chekur} and
called, for brevity sake, the semimicroscopical approach. Another aim 
is to analyze
the strength functions of the inverse resonances ($SDR^{(+)}$) in
\nuc{208}{Tl}
within the same approach. The basic points of the semimicroscopical
approach
are the following: (i) continuum-RPA (CRPA) calculation of the reaction
amplitu\-des corresponding to the excitation of the considered GR by an
external single-particle field; (ii) Breit-Wigner parameterization of the
calculated amplitudes to deduce the parameters of those
particle-hole-type 
doorway states, which form the GR; (iii) a phenomenological description
(with averaging over the energy) of the doorway-state coupling to
many-quasiparticle configurations. In the considered case the only
adjustable parameter needed for the description is found by comparison
of
the calculated and experimental total widths of the $SDR^{(-)}$. The
phenomenological mean field and the isovector part of the Landau-Migdal
particle-hole interaction are used in the calculations. The calculation
results
are compared with both the experimental data taken from
Ref.~\cite{akimun}
and results of some previous calculations of the $GTR$ partial proton
widths~\cite{chekmurur}.

\section{Calculation scheme}


\noindent{\it CRPA equations.}\\
All CRPA-equations used in this work are given in the form accepted 
within the finite Fermi-system theory~\cite{mig}. Let $\hat
V_{ {JLSM}}^{(\mp)}=\sum\limits_{ {a}}V_{ {JLSM}}^{(\mp)}(x_{ {a}})$, 
$V_{ {JLSM}}^{(\mp)}(x_{ {a}})=V(r)T_{ {JLSM}}(\pol n)\tau ^{(\mp )}$
be an external single-particle field acting upon the nucleus in the 
process of GR excitation. Here, $T_{ {JLSM}}(\pol n)$ is the irreducible 
spin-angular tensor operator of rank $J$, $\tau^{(\mp )},\tau ^{ {(3)}}$ 
are the Pauli matrices acting in the isospin space. Bearing in mind the 
$SDR^{(\mp)}$ excitation we put below $J=0,1,2;~L=S=1$  (for the $GTR$ 
$J=S=1, L=0$). Within the CRPA the strength function (''inclusive 
reaction'' cross section) is defined by the following expression:

\begin{equation}
S_{ {V,J}}^{(\mp )}(\omega )=-{\frac 1\pi }\im P_{ {V,J}}^{(\mp )}
(\omega )=-{\frac 1\pi }\im\int V(r)A_{ {J}}^{(\mp )}(r,r^{\prime };
\omega )\tilde V_{ {J}}^{(\mp)}(r^{\prime },\omega ){\d} r{\d} r',
\label{eq1}
\end{equation}

\noindent    where $P_{ {V,J}}^{(\mp)}(\omega)$ is the nuclear 
polarizability (''forward scattering'' amplitude) corresponding to the 
given external field, 
$(rr^{\prime })^{ {-2}}A_{ {J}}^{(\mp)}(r,r^{\prime };\omega )$ 
is the radial part of free particle-hole propagator carrying quantum 
numbers $J,~L$ and $S$,  $\omega $ is the excitation energy measured from 
the parent-nucleus ground-state energy and $\tilde V_{ {J}}^{(\mp )}$ 
are the so-called effective fields. They satisfy the integral equations:

\begin{equation}
\tilde V_{ {J}}^{(\mp )}(r,\omega )=V(r)+{\frac{2G^{\prime }}
{{r^{ {2}}}}}\int A^{(\mp)}_{ {J}}(r,r^{\prime };\omega )\tilde 
V_{ {J}}^{(\mp )}(r',\omega){\d} r',  
\label{eq2}
\end{equation}

\noindent  where $G'$ is the intensity of the spin-isospin part of
Landau-Migdal particle-hole interaction 
$(F'+G'\pol\sigma_1\pol\sigma_2)(\pol\tau_1\pol\tau_2)
\cdot{\mathrm\delta}
({\pol r}_{ {1}}-{\pol r}_{ {2}})$. Propagators $A_{ {J}}^{(\mp )}$ 
can be expressed in terms of: (i) occupation numbers $n_{ {\mu}} 
^{ {\alpha}} $ ($ {\alpha =n,p}$), (ii) radial bound-state
single-particle wave functions $r^{ {-1}}\chi_{ {\mu}}^{ {\alpha}}(r)$
 ($ {\mu=\varepsilon _\mu ,j_\mu ,l_\mu} $) and  Green functions 
$g_{ {(\nu )}}^{ {\alpha}}(r,r^{\prime };\varepsilon )$ of the radial 
single-particle Schrodinger equations ($ {(\nu )=j_\nu ,l_\nu }$). The 
expressions for $A_{ {J}}^{(\mp)}$ are well-known and given, for 
instance, in Ref.~\cite{murur} for the case of the $GTR$ description.

The alternative representation of $S_{ {V,J}}^{(\mp)}$, which is
more convenient for considera\-tion of continuum problems, follows from
Eqs.~(\ref{eq1}),~(\ref{eq2}) (below superscripts $(\mp)$ are sometimes
omitted):
\begin{equation}
S_{ {V,J}}(\omega )=-{\frac 1\pi }\im\int \tilde V_{ {J}}^{*}(r,\omega
)A_{ {J}}(r,r^{\prime };\omega )\tilde V_{{J}}(r^{\prime },\omega ){\d}r
{\d} r'=\sum\limits_{ {\mu,(\nu)}}|M_{ {c}}(\omega )|^{ {2}}.
\label{eq3}
\end{equation}
The expres\-sion for the "reaction-amplitudes" $M^{(-)}_{ {c}}$ has the
form:
\begin{equation}
M_{ {c}}^{(-)}
=(n_{ {\mu}} ^{ {n}})^{ {1/2}}t_{ {(\nu )(\mu )}}^{ {JLS}}\int 
\chi _{ {\varepsilon (\nu)}}^{ {p}}(r)\tilde V_{ {J}}^{(-)}(r,\omega )
\chi _{ {\mu}}^{ {n}}(r){\d} r.  \label{eq4}
\end{equation}
Here, $t_{ {(\nu )(\mu )}}^{ {JLS}}=(2J+1)^{ {-1/2}}\left\langle (\nu)
||T_{ {JLS}}||(\mu )\right\rangle $ is the kinematic factor, 
$r^{ {-1}}\chi_{ {\varepsilon (\nu )}}^{ {p}}(r)$ is the normalized to 
the $\mathrm{\delta}$ - function of energy radial continuum-state (real)
wave function for the escaping proton
with energy $\varepsilon =\omega +\varepsilon _{ {\mu}} ^{ {n}}$,
$ {c=J,L,S,(\nu),\mu }$ could be considered as a set the 
reaction-channel quantum numbers. A similar expression takes place for 
amplitudes $M_{ {c}}^{(+)}{}$.


\noindent{\it Doorway-state parameters.}\\
In the vicinity of the GR with not-too-large excitation energy the 
reaction amplitudes calculated within the CRPA exhibit narrow, 
as a rule nonoverlapping, resonances. These resonances correspond to the 
particle-hole-type doorway states forming the GR. Breit-Wigner 
parameterization of the ampli\-tudes $P_{ {V,J}}$ and $M_{ {c}}$

\begin{equation}
P_{ {V,J}}(\omega )=\sum\limits_{ {g}}{\frac{R_{ {g}}}{{\omega
-\omega _{ {g}}+{\frac {\mathrm{i}}2}\Gamma _{ {g}}^{\uparrow }}}}\ ;
\ M_{ {c}}(\omega )=\frac 1{\sqrt{2\pi }}\e^{{\mathrm{i}}\xi_c(\omega )}
\sum\limits_{ {g}}{\frac{R_{ {g}}^{ {1/2}}\left(\Gamma_{ {gc}}^{\uparrow
}\right) ^{ {1/2}}}{{\omega -\omega _{ {g}}+{\frac {\mathrm{i}}2}
\Gamma_{ {g}}^{\uparrow }}}}\
\label{eq5}
\end{equation}

\noindent allows one to deduce the doorway-state parameters: energy 
$\omega_{ {g}}$, partial strength $R_{ {g}}$, partial and total escape 
widths $\Gamma_{ {gc}}^{\uparrow }$ and $\Gamma _{ {g}}^{\uparrow }$, 
respectively. The possibility to use the above parameterization can be 
checked by satisfying the equality $\Gamma _{ {g}}^{\uparrow }=\sum
\limits_{ {\mu,(\nu)}}\Gamma _{ {gc}}^{\uparrow }$, which follows from 
Eqs.~(\ref{eq1}),~(\ref{eq3}), (\ref{eq5}). The phases
$\xi _{ {c}}(\omega )$ in Eq.~(\ref{eq5}) are
smooth functions of energy. Note that amplitudes $R_{g}^{ {1/2}},(\Gamma
_{ {gc}}^{\uparrow })^{ {1/2}}$ are not sign-fixed quantities and only 
their products found with the help of parameterization (\ref{eq5}) are 
used below.

The integrals $R_{ {V,J}}^{(\mp )}=\int\limits_{\omega_{ {1}}}^
{\omega_{ {2}}}S_{ {V,J}}^{(\mp)}(\omega){\d}\omega$ calculated with the
use of Eq.~(\ref{eq3}) for a rather wide excitation energy interval
$\delta_{ {12}}=\omega_{ {2}}-\omega_{ {1}}$ define the GR total
strength, so that ratios $x_{ {g}}=R_{g}/R_{ {V,J}}$ are the
relative partial strengths. Calculations of the above strength functions
can be checked by a comparison of difference 
$R_{ {V,J}}^{(-)}-R_{ {V,J}}^{(+)}$ and only slightly model-dependent 
sum rule $(SR)_{ {V}}=\int \rho^{(-)}V^{ {2}}(r)r^{ {2}}{\d} r$, where
$\rho ^{(-)}=\rho ^{ {n}}-\rho ^{ {p}}$ is the neutron-excess density. 
Hence the ratios $x_{ {V,J}}=(R^{(-)}_{ {V,J}}-R^{(+)}_{ {V,J}})/
(SR)_{ {V}}$ of above quantities should be close to unity provided that
interval $\delta_{ {12}}$ is sufficiently large. In the case of $GTR$
and $SDR$ ratios $B_{ {V,J}}=R_{ {V,J}}^{(+)}/R_{ {V,J}}^{(-)}$ 
calculated for a long-wave external field $V(r)$ are expected to be 
small for nuclei with large neutron excess due to Pauli blocking.


\noindent{\it Doorway-state coupling to many-quasiparticle
configu\-ra\-tions.}\\
This coupling leads to doorway-state spreading and formation of the GR 
as a single resonance in the energy dependence of energy-averaged 
reaction cross sections. We take this coupling into consideration 
phenomenologically by independently spreading  each doorway-state 
resonance~\cite{murur,chekur}. It means that the transition to the
energy-averaged reaction amplitudes $\bar P_{ {V,J}}$ and 
$\bar M_{ {c}}$ can be realized by the following substitution in
Eqs.~(\ref{eq5}): $\omega -\omega _{ {g}}+{\frac { \mathrm{i}}2}
\Gamma _{ {g}}^{\uparrow }\to \omega-\omega _{ {g}}+{\frac 
{ \mathrm{i}}2}\Gamma _{ {g}}^{}$, 
where $\Gamma _{ {g}}=\Gamma _{ {g}}^{\uparrow}+\Gamma ^{\downarrow }$.
The doorway-state spreading width $\Gamma^{\downarrow }$ is considered
as the only adjustable parameter of the semimicroscopical approach. It 
can be found by equating the total width $\Gamma$ (dependent on 
$\Gamma ^{\downarrow }$) of the calculated energy-averaged strength
function of the $SDR^{(-)}$
\begin{equation}
\bar S_{ {V}}^{(-)}(\omega )=\sum\limits_{ {J=0,1,2}}(2J+1)\bar 
S_{ {V,J}}^{(-)}(\omega)\ ;\ \bar S_{ {V,J}}^{(-)}(\omega )=
-{\frac 1\pi }\im\sum\limits_{ {g}}{\frac{R_{g}}{{\omega-\omega _{ {g}}
+{\frac { \mathrm{i}}2}\Gamma _{ {g}}}}}
\label{eq6}
\end{equation}
to the total width $\Gamma ^{\exp}$ of the $SDR^{(-)}$ in the
experimental inclusive reaction cross section. Because this cross 
section is parameterized by a single-level formula~\cite{akimun},
we approximate calculated strength function~(\ref{eq6}) by the same
formula:
\begin{equation}
\bar S_{ {V}}^{(-)}\to\frac 1{2\pi }\cdot \frac{R_{ {V}}^{(-)}\Gamma }
{(\omega -\omega_{ {m}})^{ {2}}+\frac 14\Gamma ^{ {2}}},  
\label{eq7}
\end{equation}
where $R_{ {V}}^{(-)}=\int\limits_{\omega_{ {1}}}^{\omega_{ {2}}}
S_{ {V}}^{(-)}(\omega ){\d}\omega $ and $\omega _{ {m}}$ are,
respectively, the calculated total strength and mean excitation 
energy of the $SDR^{(-)}$.

Because each doorway-state resonance in the energy dependence of 
amplitudes $\bar M_{ {c}}(\omega )$ becomes rather broad, it is 
necessary to take also into account changing   the penetrability 
of the potential barrier for escaping protons over the resonance. 
It can be done as follows~\cite{chekmurur}:
\begin{equation}
|\bar M_{c}(\omega )|^2=\frac 1{2\pi}
\left| \sum\limits_{ {g}}{\frac{R_{g}^
{ {1/2}}\left( \bar \Gamma_{ {gc}}^{\uparrow }\right) ^{ {1/2}}}
{{\omega -\omega _{ {g}}+{\frac {\mathrm{i}}2}\Gamma _{ {g}}}}}\right|
^{ {2}}\ ;\ \bar \Gamma _{ {gc}}^{\uparrow }=\Gamma _{ {gc}}
^{\uparrow }\frac{\bar P_{ {g(\nu )}}}{P_{ {(\nu )}}(\varepsilon 
_{ {g\mu }})}.  
\label{eq8}
\end{equation}
Here, $\varepsilon _{ {g\mu }}=\omega _{ {g}}+\varepsilon^{ {n}}_{ {\mu}} 
,\bar P_{ {g(\nu )}}$ is the penetrability averaged over the resonance:
\begin{equation}
\bar P_{ {g(\nu )}}=\frac 1{\sqrt{2\pi }\sigma _{ {g}}}\int 
{P_{ {(\nu )}}(\varepsilon )}\exp \left( -\frac{(\varepsilon 
-\varepsilon _{ {g\mu }})^{ {2}}}{2\sigma _{ {g}}^{ {2}}}
\right) {\d}\varepsilon \ ;\ \sigma _{ {g}}=\Gamma _{ {g}}/2.35.
\label{eq9}
\end{equation}

Thus, the energy-averaged partial cross sections $\bar \sigma _{ {\mu}}
(\omega )=\sum\limits_{ {(\nu )}}|\bar M_{ {c}}(\omega )|^{ {2}}$ (the 
fluctuational part of these cross sections is neglected) can be 
calculated without the use of any free parameters. Summation in the 
above equation is performed over the quantum numbers of the escaping proton, 
which are compatible with the selection rules for the spin-dipole 
transitions. Cross section $\overline{\sigma }_{ {\mu}} (\omega )$ 
corresponds to population of single-hole state ${ {\mu}} ^{ {-1}}$ 
in the product nucleus after the $SDR^{(-)}$ proton decay.


\noindent{\it Branching ratios.}\\ 
The $SDR^{(-)}$ branching ratios and partial 
widths for the direct proton decay are defined as
follows:
\begin{equation}
b_{ {\mu}} =\left. \int \sum\limits_{ {J=0,1,2}}(2J+1)\bar 
\sigma _{ {\mu}} ^{ {J}}(\omega){\d}\omega \right/ \int \bar
S_{ {V}}^{(-)}(\omega ){\d}\omega \ ,
\Gamma^{\uparrow}_{ {\mu}}=b_{ {\mu}}\Gamma,  
\label{eq10}
\end{equation}
where $\bar S_{ {V}}^{(-)}$ is defined by Eqs.~(\ref{eq6}). Note that
this strength function, its single-level approximation (\ref{eq7}) 
(parameters $R_{ {V}}^{(-)}$ and $\omega_{ {m}}$), partial branching 
ratios determined by Eqs.~(\ref{eq8})--(\ref{eq10}) are somewhat
dependent on the energy interval $\delta_{ {12}}$ used in the CRPA
analysis. The choice of the interval is connected with description of
corresponding experimental data (see next Section).

When only one doorway state corresponds to the considered GR (for 
instance, in case of the $GTR$) one can calculate the average partial 
escape widths of this GR directly with the help of Eq.~(\ref{eq8}):
$\bar \Gamma _{ {\mu}} ^{\uparrow}=\sum\limits_{ {(\nu )}}\bar 
\Gamma _{ {c}}^{\uparrow }$. Such a procedure was realized in 
Ref.~\cite{chekmurur}. To take the averaged potential-barrier
penetrability more accurately into consideration in accordance with 
Eqs.~(\ref{eq8}),~(\ref{eq9}) we use the experimental one-hole state 
energies $\varepsilon^{ {n}\exp}_{ {\mu}}$ instead of calculated ones. 
For comparison with experimental data the calculated quantities 
$b_{ {\mu}} $ or $\bar \Gamma _{ {\mu}}^{\uparrow }$ should be also 
multiplied by spectroscopic factors $S_{ {\mu}} $ of the corresponding 
single-hole states in the product nucleus \nuc{207}{Pb}.

\section{Calculation results}


\noindent{\it Choice of model parameters.}\\
The nuclear mean field and particle-hole interaction are the input 
data for any RPA calculations. In the following the isoscalar part 
of the phenomenological nuclear mean field (including the spin-orbit 
interaction) is chosen in accordance with Ref.~\cite{murur}. Only
mean field amplitude $U_{ {0}}$ is slightly increased (54.1~MeV instead
of 53.3~MeV) to describe better the nucleon separation energies for
\nuc{208}{Pb}. The strengths $(F',G')=(f',g')\cdot300~{ {\mathrm{MeV}
~\cdot \mathrm{fm}^{ {3}}}}$ of the isovector part of the Landau-Migdal
particle-hole interaction are chosen as follows: $f'=1.0$ (to describe
the experimental difference of the neutron and proton separation
energies in \nuc{208}{Pb}); $g'=0.76$ (to describe the experimental
energy of the $GTR$ in \nuc{208}{Bi}~\cite{akimun}). The above strengths
are close to those used in Refs.~\cite{murur,chekmurur}. The isovector
part $\frac 12\tau ^{ {(3)}}v(r)$ of the mean field is calculated
in a  self-consistent way (see e.g. Ref.~\cite{rumur}):
$v(r)=2F'\varrho^{(-)}$.
The nuclear Coulomb field is calculated in Hartree
approximation via the proton density $\varrho ^{ {p}}$.


\noindent{\it Calculation results for the \nuc{208}{Pb} 
parent nucleus.}\\
The nuclear mean field chosen above allows one to describe satisfactorily 
the single-quasiparticle spectrum of the \nuc{208}{Pb} parent nucleus. 
The calculated neutron single-hole spectrum
(energies $\varepsilon_{ {\mu}}^{ {n}} $) is given in the 
Table~\ref{tab1} in comparison with the experimental one (energies 
$\varepsilon _{ {\mu}}^{ {n}\exp}$). All experimental quantities (except
for the spectroscopic factors $S_{ {\mu}}$ and neutron separation energy
$S_{ {n}}$) in this Table are taken from Ref.~\cite{akimun}. The values
of $S_{ {\mu}}$ and $S_{ {n}}$ are taken from Refs.~\cite{specfac}
and \cite{wapstra}, respectively.

In CRPA calculations of the $SDR^{(-)}$ and the $GTR$ strength
functions the radial dependence of the external field $V(r)$ is
chosen as $r/R$ and $1$, respectively ($R$ is the nuclear radius).
The $GTR$ strength function is similar to that given in~\cite{murur}
and is not shown here. The $SDR^{(-)}$ strength functions 
$S^{(-)}_{ {V,J}}(\omega^{(-)})$ are shown in Figs.~1a--1c up
to the excitation energy $\omega^{(-)}_{ {2}}=35\:  \mathrm{MeV}$
($\omega^{(-)}=\omega-\Delta^{(-)}_{\exp}$ is the excitation energy 
measured from the \nuc{208}{Bi} ground-state energy, $\Delta^{(-)}
_{\exp}=3.66\: \mathrm{MeV}~\cite{wapstra}, \Delta^{(-)}
_{ \mathrm{calc}}=3.30\: \mathrm{MeV}$).

In the case of $J^{ {\pi}} =0^{-}$ and $J^{ {\pi}} =1^{-}$ the main 
part of the total strength is exhausted by one doorway state (85\%
and 81\%, respectively). In the case of $J^{ {\pi}} =2^{-}$ the 
calculated strength function exhibits an essential gross structure: 
eight doorway-state resonances exhaust 91\% of the total strength 
(the relative strength of most resonances $x_{ {g}}$ (in \%) is shown 
in Figs.~1a-1c). The ratios $x_{ {J}}$ and $B_{ {J}}$ are also given 
in these figures.

The energy-averaged strength
functions $\bar S_{ {V}}^{(-)}(\omega^{(-)})$ calculated according to equations 
(\ref{eq6}) for three excitation-energy 
intervals $\delta_{12}=\omega^{(-)}_{ {2}}-\omega^{(-)}_{ {1}}$  
($\omega^{(-)}_{ {2}}=35\: \mathrm{MeV}$)
are shown in Fig.~2. The chosen 
energies $\omega^{(-)}_{ {1}}$ and the  relative $SDR^{(-)}$ total 
strength $x_{ {\delta}}$ are given in Table~\ref{tab2}. The strength
function calculated for each interval is approximated by single-level 
formula~(\ref{eq7}) so that adjustable parameter $\Gamma^{\downarrow}$
(doorway-state spreading width) and calculated $SDR^{(-)}$ energy 
$\omega^{(-)}_{ {m}}$ are determined by fitting to the $SDR^{(-)}$
experimental total width $\Gamma^{\exp}=8.4\:  \mathrm{MeV}$
~\cite{akimun}. These parameters are also given in
Table \ref{tab2}.

Partial and total branching ratios for the $SDR^{(-)}$ proton decay are
calculated according to equations (\ref{eq8})--(\ref{eq10}). Calculated
total branching ratios $b$ depending on considered interval $\delta
_{ {12}}$ are given in Table \ref{tab2}. The use of
$\omega _{ {1}}^{(-)}=17{}\; \mathrm{MeV}$ seems to be reasonable for
description of experimental data~\cite{akimun}. The reasons are the
following: (i) calculated strength function $\bar S_{ {V}}^{(-)}(\omega
^{(-)})$ is satisfactorily described by single-level formula 
(see Fig.~2) used for approximation of the experimental inclusive 
reaction  cross section~\cite{akimun}; (ii) the most part of the
calculated  $SDR^{(-)}$ strength ($x_{ {\delta}} =83\%$) is exhausted
within this interval $\delta _{ {12}}$; (iii) doorway-state spreading
width $\Gamma^{\downarrow }=4.7\; \mathrm{MeV}$ found with the use of
the experimental $SDR^{(-)}$ total width is reasonably greater than
the experimental $GTR$ spreading width $\Gamma_{ {GTR}}^{\downarrow }
=3.54\: \mathrm{MeV}$~\cite{akimun}. 
The partial $SDR^{(-)}$ proton branching ratios 
$b_{ {\mu}} $ ($b=16.1\%$) calculated for the chosen
excitation-energy interval   are given in Table~\ref{tab1}.

The averaged proton 
partial widths of the $GTR${} $\bar \Gamma_{ {\mu}}$ calculated 
according to Eqs.~(\ref{eq8}),~(\ref{eq9}) 
are also given in
Table~\ref{tab1} in comparison with the experimental data.  In above
calculations the following expression for the penetrability is used 
~\cite{bor}: $P(kR)=kR\left[F^{ {2}}(kR)+G^{ {2}}(kR)\right] ^{ {-1}}$,
where $k^{ {2}}=2m\varepsilon /\hbar ^{ {2}}$, $F$ and $G$ are 
well-known Coulomb functions.

The strength functions of the second $SDR^{(+)}$ are also calculated 
within CRPA for the external field $V(r)=r^{ {3}}/R^{ {3}}$. The mean 
excitation energy $\omega _{ {m}}^{(+)}$ of this GR is found to be about 
19.2 MeV ($\omega^{(+)}=\omega -\Delta ^{(+)}_{\exp}$ is the excitation 
energy measured from the \nuc{208}{Tl} ground-state energy, $\Delta
^{(+)}_{\exp}=4.21\: \mathrm{MeV}$~\cite{wapstra}, $\Delta^{(+)}
_{ \mathrm{calc}}=3.80\: \mathrm{MeV}$).


\noindent {\it Discussion of results.}\\
The quality of the performed CRPA calculations of the $SDR^{(-)}$ 
spin-dipole strength functions is satisfactory, because calculated 
ratios $x_{ {V,J}}$ are close to unity (Figs.~1a--1c). As expected,
Pauli blocking leads to suppression of the $SDR^{(+)}$ spin-dipole 
strengths $R^{(+)}_{ {J}}=B_{ {J}}(1-B_{ {J}})^{ {-1}}x_{ {J}}(SR)
_{ {V}}${} ($(SR)_{ {V}}=2.38$ for \nuc{208}{Pb}), because
calculated ratios $B_{ {J}}$ are rather small (Figs.~1a--1c).

Calculated proton partial widths of the $GTR$ are in a reasonable 
agreement with both the experimental data and the results of previous 
calculations performed within the same approach. The difference 
between two sets of $\overline{\Gamma }_{ {\mu}} ^{\uparrow }$ is
explained by the self-consistent calculation of the mean Coulomb
field, by the small difference of model parameters used in 
calculation  and by the use in this work of spectroscopic factors 
$S_{ {\mu}}$ for final states in \nuc{207}{Pb}.

The calculated mean excitation energy of the second $SDR^{(+)}$ in 
\nuc{208}{Tl} $\omega^{(+)}=19.2\: \mathrm{MeV}$ can be considered
as a guide for an experimental search of this GR.

The calculated mean $SDR^{(-)}$ energy $\omega^{(-)}_{ {m}}=23.1\: 
\mathrm{MeV}$ is in acceptable agreement with experimental energy
$\omega _{\exp}^{(-)}=21.1\pm0.8\: \mathrm{MeV}$~\cite{akimun}.
However, the main result of this work consists in a reasonable 
description of the $SDR^{(-)}$ total branching ratio ($b=16.1\%,
b^{\exp }=14.1\pm 4.2\%$~\cite{akimun}) or of the $SDR^{(-)}$ total
proton width ($\overline{\Gamma }_{ \mathrm{tot}}^{\uparrow }=1.35\:
\mathrm{MeV},\overline{\Gamma }_{ \mathrm{tot}}^{\uparrow \exp }
=1.18\pm 0.35\: \mathrm{MeV}$~\cite{akimun}).

\section{Conclusion and acknowledgments}

In this work the semimicroscopical approach is applied to calculate the
branching ratios for the direct proton decay of the $SDR^{(-)}$ in 
\nuc{208}{Bi}. A reasonable description of the experimental data on the 
proton total branching ratio is obtained. Previous calculations on the 
partial proton widths of the $GTR$ in \nuc{208}{Bi} are refined. The 
energy position of the second $SDR^{(+)}$ in \nuc{208}{Tl} is predicted.

It will be possible to perform more detailed comparison of the results
obtained within the semimicroscopical approach and experimental data
provided that the experimental partial proton branching ratios 
$b_{ {\mu}} $ will become available and different $SDR^{(-)}$ spin 
components will be separated~\cite{har}. In this connection further
experimental and theoretical studies of proton and $\gamma $ - decays of
the $SDR^{(-)}$ seem to be very promising.

The authors are grateful to Prof. M.N. Harakeh for stimulating discussions
and valuable remarks. One of the authors (M.H.U.) acknowledges generous
financial support from the ``Nederlandse organisatie voor wetenschappelijk 
onderzoek'' NWO during his stay at the KVI.


\phantom{givi}
\phantom{givi}
\phantom{givi}
\phantom{givi}
\phantom{givi}

\pagebreak

\begin{center}
\includegraphics[width=15.cm]{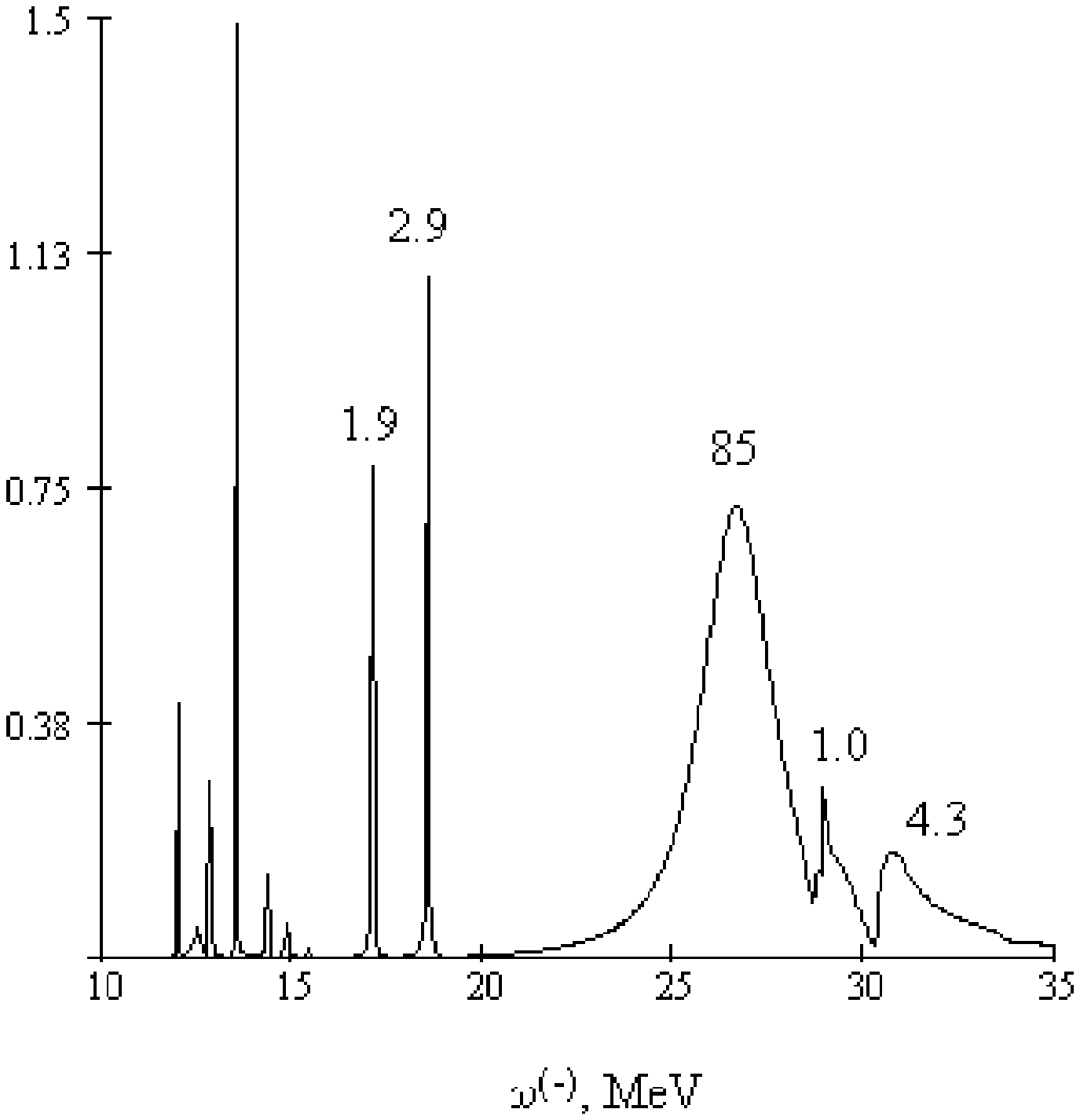}
\end{center}
{{\bf Figure 1a:} Spin-dipole strength function calculated within
CRPA for the $J^\pi=0^-$- component of the $SDR^{(-)}$ in $^{208}$Bi.
The relative strengths (in \%) of doorway states are also shown.
x=0.98, B=14\%.}
\pagebreak

\begin{center}
\includegraphics[width=15.cm]{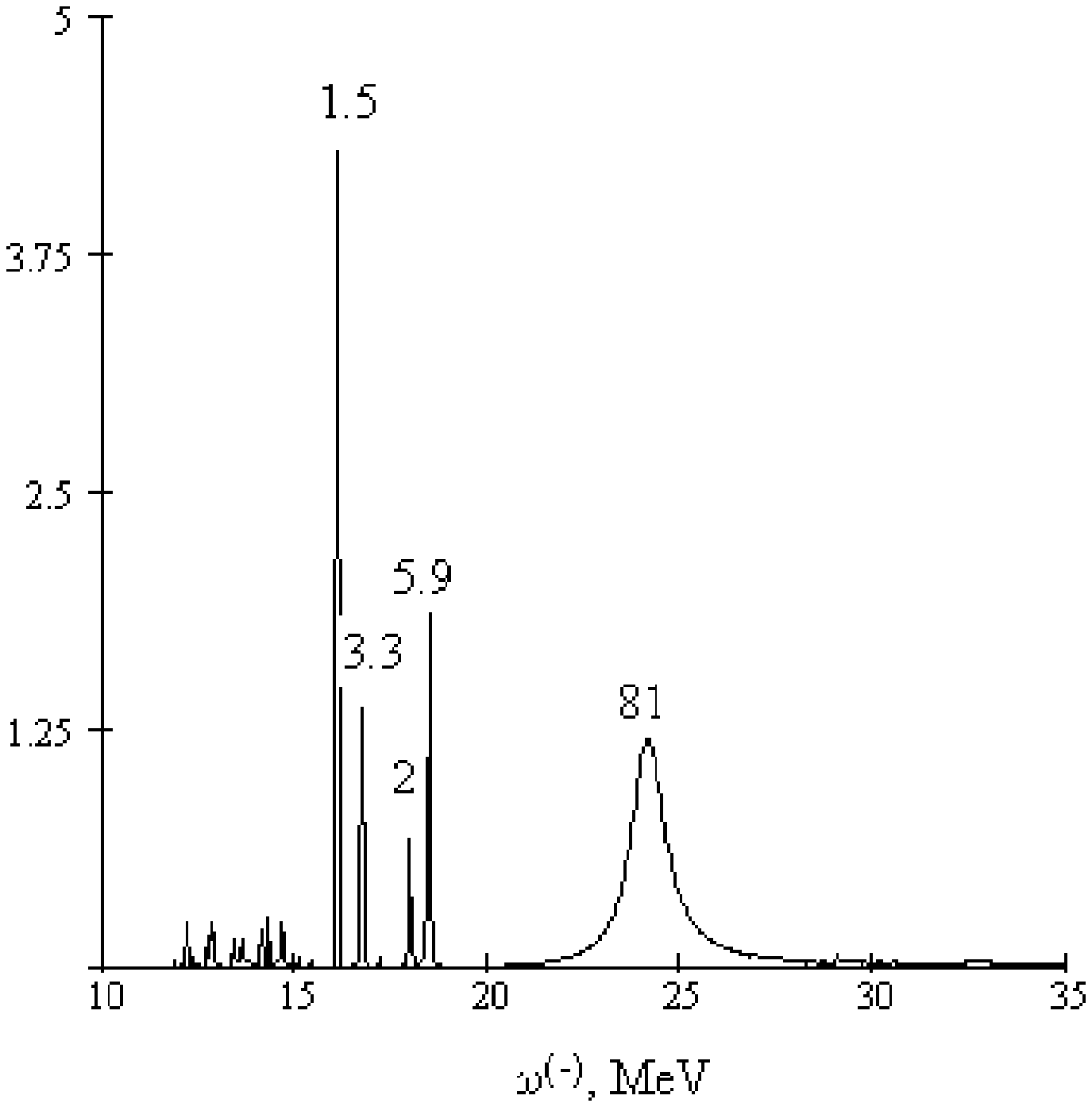}
\end{center}
{{\bf Figure 1b:} The same as in {\bf Fig. 1a}, but for $J^\pi=1^-$.
x=0.98, B=8\%.}
\pagebreak

\begin{center}
\includegraphics[width=15.cm]{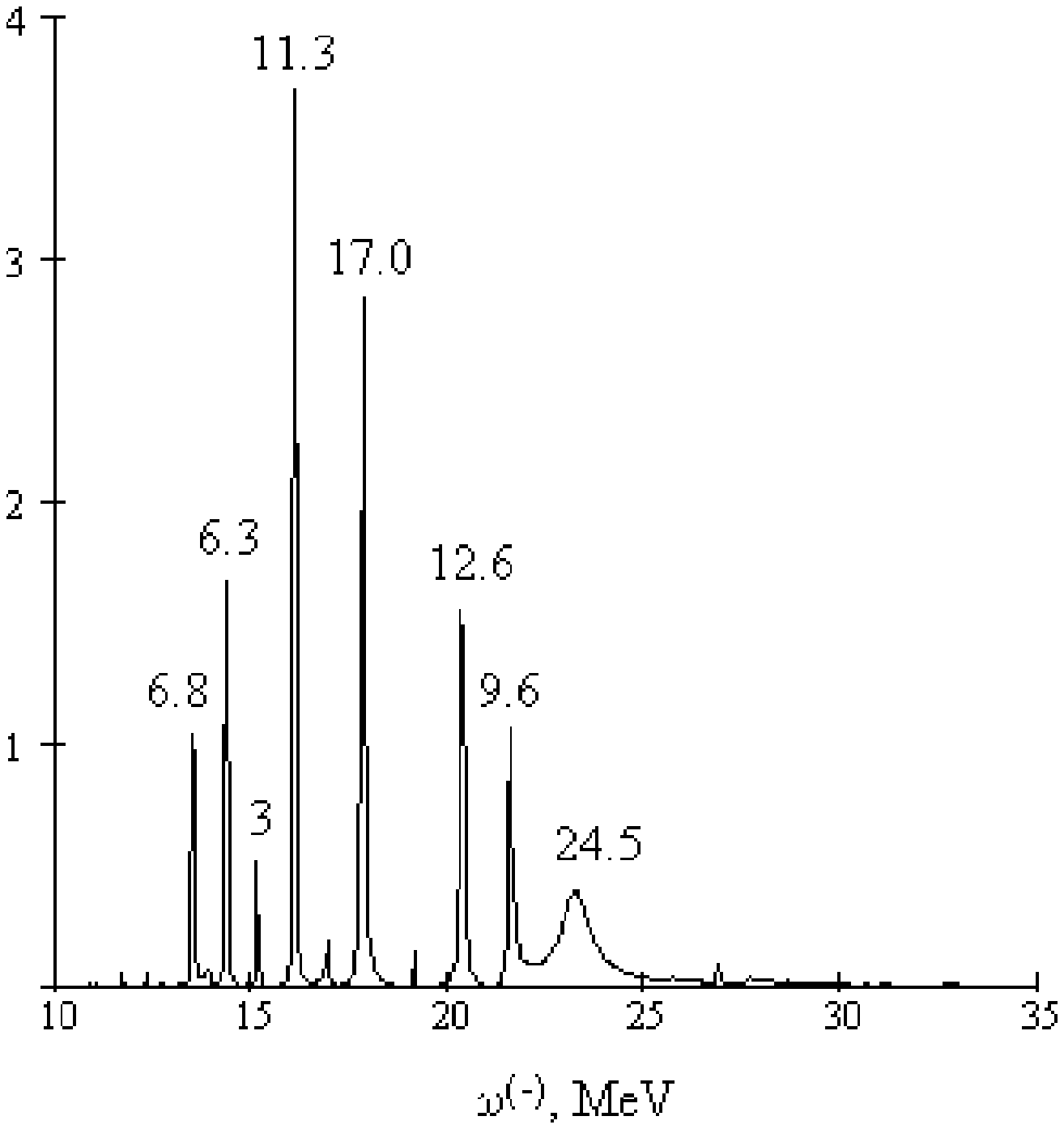}
\end{center}
{{\bf Figure 1c:} The same as in {\bf Fig. 1a}, but for $J^\pi=2^-$.
x=0.99, B=3.5\%.}

\pagebreak
\begin{center}
\includegraphics[width=15.cm]{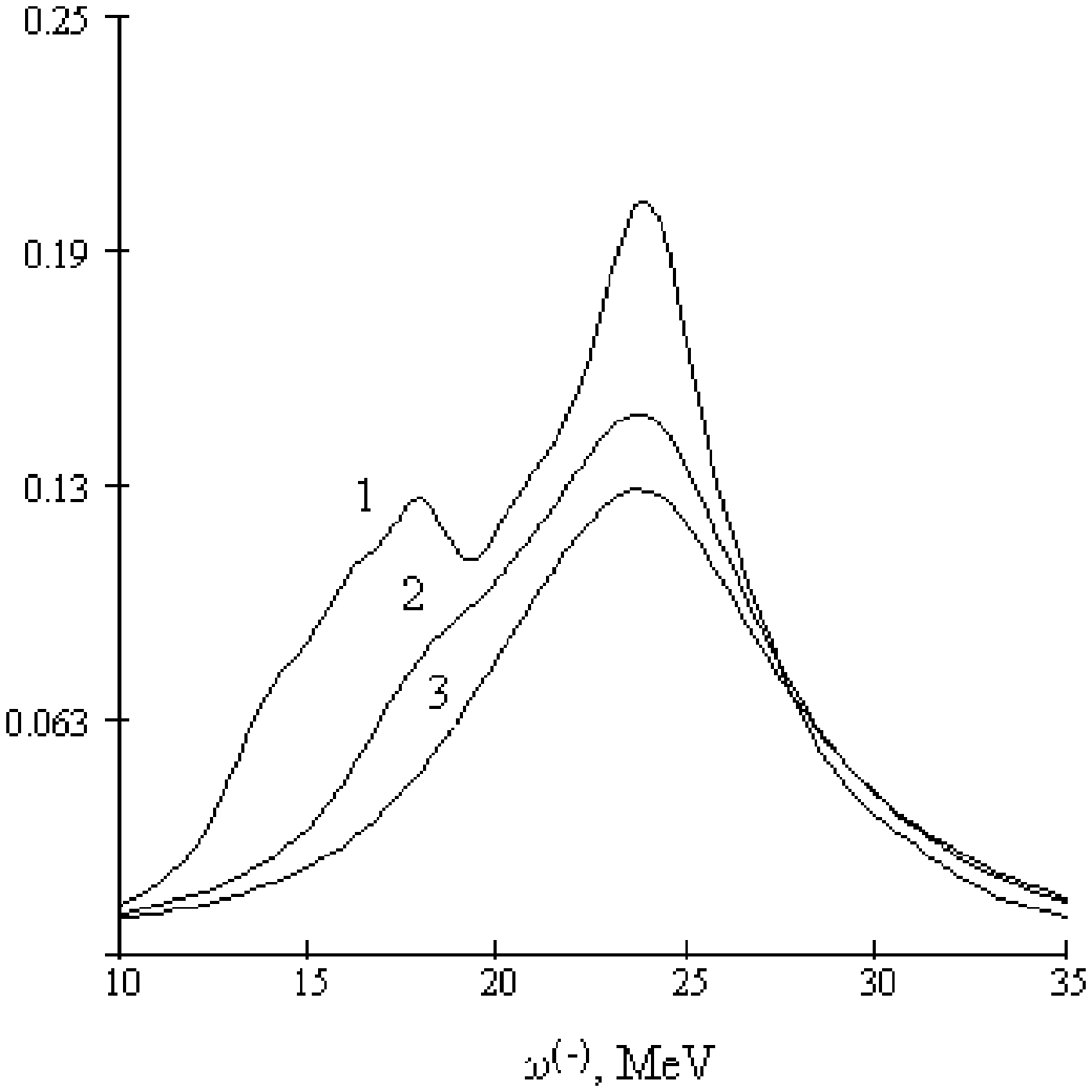}
\end{center}
{{\bf Figure 2:} Energy-averaged strength functions $S^{(-)}_V
(\omega^{(-)})$
calculated for different intervals $\delta_{12}$: $\omega^{(-)}_1$=13,
17 and 18 MeV (curves 1, 2 and 3, respectively).}

\phantom{givi}
\pagebreak
\begin{table}[t]
\caption{Calculated escape widths and branching ratios for proton decay
of the $GTR$ and the $SDR^{(-)}$ in \nuc{208}{Bi}. Rather small
contribution of deep-hole states to the calculated $b$ value
($\sum_{\mathrm{\mu'}}b_{\mathrm{\mu'}}=1.94\%$) is not shown.
Experimental data contain
also $\Gamma^{\uparrow}_{\mathrm{tot}}=1.18\pm0.35\:\mathrm{MeV}$
and $b=14.1\pm4.2\%$ for the $SDR^{(-)}$.}

\label{tab1}

\begin{center}
\begin{tabular}{|rcc|cccccc|}
\hline
\multicolumn{1}{|r|}{} & \multicolumn{1}{c|}{} & \multicolumn{1}{c|}{} &
\multicolumn{1}{|c|}{} & \multicolumn{3}{c|}{$GTR$} & \multicolumn{2}{|c|}{$
SDR^{(-)}$} \\ \cline{5-9}
\multicolumn{1}{|r|}{
\begin{tabular}{c}
$\mu ^{{\rm -1}}$ \\
\phantom{gi} \\
\phantom{go}
\end{tabular}
} & \multicolumn{1}{c|}{
\begin{tabular}{c}
$S_{{\rm \mu }}$ \\
\phantom{gi} \\
{\footnotesize Ref. \cite{specfac}}
\end{tabular}
} & \multicolumn{1}{c|}{${\rm
\begin{tabular}{c}
$-\varepsilon _{\mu ,}^n$ \\
${\rm MeV}$ \\
\phantom{go}
\end{tabular}
}$} & \multicolumn{1}{|c|}{${\rm
\begin{tabular}{c}
$-\varepsilon _{\mu ,}^{n\exp }$ \\
${\rm MeV}$ \\
\phantom{go}
\end{tabular}
}$} & \multicolumn{1}{c|}{$
\begin{tabular}{c}
$S_{{\rm \mu }}^{{\rm -1}}\overline{\Gamma }_{{\rm \mu }}^{\uparrow },$ \\
${\rm keV}$ \\
{\footnotesize Ref. \cite{chekmurur}}
\end{tabular}
$} & \multicolumn{1}{c|}{$
\begin{tabular}{c}
$\overline{\Gamma }_{{\rm \mu }}^{\uparrow },$ \\
${\rm keV}$ \\
{\footnotesize this work}
\end{tabular}
$} & \multicolumn{1}{c|}{$
\begin{tabular}{c}
$\overline{\Gamma }_{{\rm \mu }}^{\uparrow },$ \\
${\rm keV}$ \\
{\footnotesize exp}
\end{tabular}
$} & \multicolumn{1}{c|}{
\begin{tabular}{c}
$\overline{\Gamma }_{{\rm \mu }}^{\uparrow },$ \\
keV \\
\phantom{gogy}
\end{tabular}
} & $
\begin{tabular}{c}
$b_{{\rm \mu }},$ \\
\% \\
\phantom{go}
\end{tabular}
$ \\ \hline
\multicolumn{1}{|c|}{3p$_{{\rm \frac 12}}$} & \multicolumn{1}{c|}{1.1} &
\multicolumn{1}{c|}{7.36} & \multicolumn{1}{|c|}{7.37} & \multicolumn{1}{c|}{
43} & \multicolumn{1}{c|}{60.8} & \multicolumn{1}{c|}{58.4$\pm $19.8} &
\multicolumn{1}{c|}{88.4} & 1.05 \\
\multicolumn{1}{|c|}{2f$_{{\rm \frac 52}}$} & \multicolumn{1}{c|}{0.98} &
\multicolumn{1}{c|}{8.14} & \multicolumn{1}{|c|}{7.94} & \multicolumn{1}{c|}{
48} & \multicolumn{1}{c|}{54.0} & \multicolumn{1}{c|}{incl. in p$_{{\rm %
\frac 32}}$} & \multicolumn{1}{c|}{163.2} & 1.94 \\
\multicolumn{1}{|c|}{3p$_{{\rm \frac 32}}$} & \multicolumn{1}{c|}{1.0} &
\multicolumn{1}{c|}{8.59} & \multicolumn{1}{|c|}{8.27} & \multicolumn{1}{c|}{
35} & \multicolumn{1}{c|}{50.5} & \multicolumn{1}{c|}{101.5$\pm $31.3} &
\multicolumn{1}{c|}{183.1} & 2.18 \\
\multicolumn{1}{|c|}{1i$_{{\rm \frac{13}2}}$} & \multicolumn{1}{c|}{0.91} &
\multicolumn{1}{c|}{10.19} & \multicolumn{1}{|c|}{9.00} &
\multicolumn{1}{c|}{0.78} & \multicolumn{1}{c|}{0.73} & \multicolumn{1}{c|}{
8.3$\pm $9.4} & \multicolumn{1}{c|}{318.8} & 3.80 \\
\multicolumn{1}{|c|}{2f$_{{\rm \frac 72}}$} & \multicolumn{1}{c|}{0.7} &
\multicolumn{1}{c|}{11.22} & \multicolumn{1}{|c|}{9.71} &
\multicolumn{1}{c|}{8.9} & \multicolumn{1}{c|}{5.95} & \multicolumn{1}{c|}{
15.6$\pm $7.6} & \multicolumn{1}{c|}{337.3} & 4.02 \\
\multicolumn{1}{|c|}{1h$_{{\rm \frac 92}}$} & \multicolumn{1}{c|}{0.61} &
\multicolumn{1}{c|}{11.49} & \multicolumn{1}{c|}{10.78} &
\multicolumn{1}{c|}{} & \multicolumn{1}{c|}{0.19} & \multicolumn{1}{c|}{} &
\multicolumn{1}{c|}{98.6} & 1.17 \\ \hline
& \multicolumn{3}{c|}{ total:\phantom{11.49toot}} & \multicolumn{1}{c|}{136} &
\multicolumn{1}{c|}{172.2} & \multicolumn{1}{c|}{184$\pm $49} &
\multicolumn{1}{c|}{1354.1} & 16.1 \\ \hline
\end{tabular}
\end{center}
\end{table}

Table should be put after the first mention in the text.

\phantom{gogy}
\pagebreak
\phantom{gogy}
\begin{table}[t]
\caption{Dependence of the $SDR^{(-)}$ parameters on the choice of
excitation-energy interval $\delta_{\mathrm{12}}=\omega^{(-)}
_{\mathrm{2}}-\omega^{(-)}_{\mathrm{1}}$ ($\omega^{(-)}_{\mathrm{2}}=35
\:\mathrm{MeV}$) in the analysis of strength functions
$S^{(-)}_{V,J}(\omega^{(-)})$.}
\par
\label{tab2}
\begin{center}
\begin{tabular}{|c|c|c|c|c|}
\hline
$\omega^{(-)}_{\mathrm{1}},[\mathrm{MeV}]$ & $x_{\delta}$ &
$\Gamma^{\downarrow},[\mathrm{MeV}]$ & $\omega^{(-)}_{m},
[\mathrm{MeV}]$ & $b,[\%]$ \\
\hline
13 & 1 & 2.5 & 22.7 & 21.2 \\
\hline
17 & 0.83 & 4.7 & 23.1 & 16.1 \\
\hline
18 & 0.73 & 5.8 & 23.6 & 15.5 \\
\hline
\end{tabular}
\end{center}
\end{table}

\vspace{16cm}
Table should be put after the first mention in the text.

\end{document}